\documentclass[sigconf]{acmart}
\AtBeginDocument{%
  }

\usepackage{xspace}
\usepackage{amsmath}
\usepackage{tikz}
\usepackage{pgfplots}
\usepackage{subcaption}
\usepackage{pgfplotstable}
\pgfplotsset{compat=1.18}
\usepackage{dblfloatfix}
\usepackage[subtle]{savetrees}
\usepackage[normalem]{ulem}

\newcommand{\QM}{{\mathbf{Q}}\xspace}
\newcommand{\xv}{{\mathbf{x}}\xspace}

\newcommand{\vol}{\operatorname{vol}}

\newcommand{\stitle}[1]{\vspace{0.6mm}\noindent{\bf #1.}}

\copyrightyear{2026}
\acmYear{2026}
\setcopyright{cc}
\setcctype{by}
\acmConference[Q-Data '26]{The third workshop on Quantum Computing and Quantum-Inspired Technology for Data-Intensive Systems and Applications}{May 31-June 05, 2026}{Bengaluru, India}
\acmBooktitle{The third workshop on Quantum Computing and Quantum-Inspired Technology for Data-Intensive Systems and Applications (Q-Data '26), May 31-June 05, 2026, Bengaluru, India}
\acmDOI{10.1145/3811628.3811836}
\acmISBN{979-8-4007-2703-0/2026/05}

\begin{document}

\title{Quantum Hypergraph Partitioning}


\author{Yiran Li}
\affiliation{%
 \institution{University of Toronto}
 \city{Toronto}
 \state{Ontario}
 \country{Canada}}
\email{one.li@utoronto.ca}

\author{Y. Batuhan Yilmaz}
\email{yildirim.yilmaz@mail.utoronto.ca}
\affiliation{%
 \institution{University of Toronto}
 \city{Toronto}
 \state{Ontario}
 \country{Canada}}
\author{Michael Silver}
\affiliation{%
  \institution{University of Toronto}
 \city{Toronto}
 \state{Ontario}
 \country{Canada}}
\email{m.silver@mail.utoronto.ca}

\author{Zachary Vernec}
\email{zachary.vernec@mail.utoronto.ca}
\affiliation{%
 \institution{University of Toronto}
 \city{Toronto}
 \state{Ontario}
 \country{Canada}}

\author{Hans-Arno Jacobsen}
\email{jacobsen@eecg.toronto.edu}
\affiliation{%
  \institution{University of Toronto}
  \city{Toronto}
  \state{Ontario}
  \country{Canada}
}


\renewcommand{\shortauthors}{Li et al.}

\begin{abstract}
Hypergraph partitioning is a fundamental optimization problem with
applications in data management and other domains involving higher-order
relations. In this paper, we study balanced hypergraph partitioning from the
perspective of quantum optimization. We formalize balanced 
$k$-way hypergraph partitioning with general hyperedge cut functions, and derive corresponding binary optimization formulations targeted at quantum optimization methods in both the two-way and multi-way settings. Our discussion highlights which cut functions admit Quadratic Unconstrained Binary Optimization~(QUBO) encodings and which instead lead to higher-order binary objectives or
rational forms. As a preliminary empirical validation, we focus on balanced
two-way partitioning with the all-or-nothing cut on 3-uniform hypergraphs,
where a direct QUBO is available, and evaluate simulated Quantum Approximate Optimization Algorithm~(QAOA) and Simulated
Annealing (SA) on small instances against exact solutions. The results show that
the formulation is effective on small hypergraphs and that the balance-penalty
weight plays a critical role in trading off cut quality and balance.
\end{abstract}




\maketitle

\section{Introduction}

\begin{figure}
    \centering
    \includegraphics[width=0.74\linewidth]{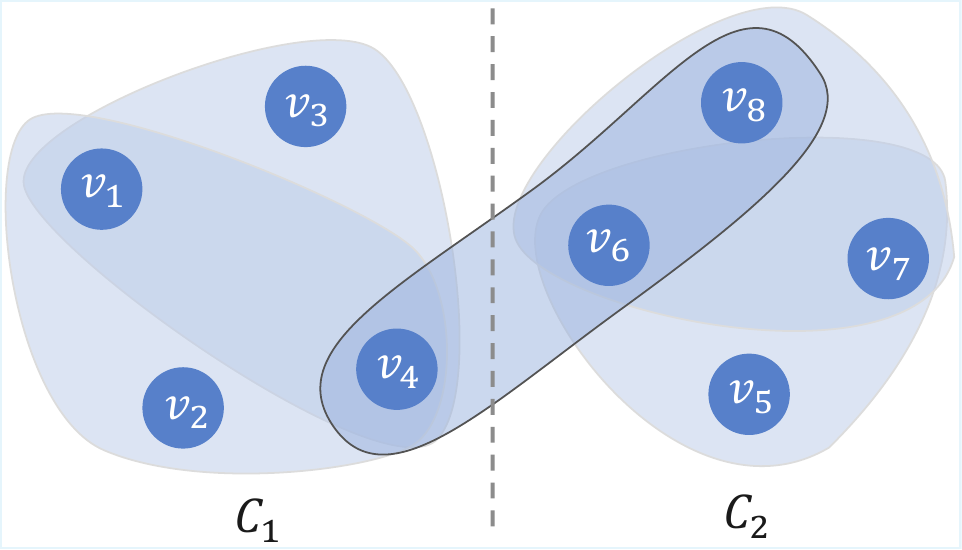}
    \vspace{-1mm}
    \caption{An example of balanced hypergraph partitioning.}
    \vspace{-3mm}
    \label{fig:balanced_hypergraph_partitioning}
\end{figure}
Hypergraphs generalize traditional graphs by allowing a hyperedge to connect more than two nodes, making them a natural model for multi-way relationships and higher-order interactions such as group chats in social networks, multi-protein interactions in biological systems, and multi-qubit gates in quantum circuits.
Hypergraph partitioning  aims to divide the nodes in a hypergraph into a designated number of subsets, such that the hyperedge connections among them are minimized. Aside from its applications to simplifying sparse-matrix multiplications~\cite{catalyurek2002hypergraph} and analyzing biological networks \cite{feng2021hypergraph}, hypergraph partitioning has been found useful for distributing quantum circuits~\cite{andres-martinezAutomatedDistributionQuantum2019}, as well as data management problems such as optimizing matrix storage layout~\cite{wuHyperMREfficientHypergraphenhanced2025} and storage sharding for distributed databases~\cite{kabiljoSocialHashPartitioner2017b}.

Figure~\ref{fig:balanced_hypergraph_partitioning} illustrates a balanced
two-way hypergraph partition, where the node set is divided into two equally
sized parts \(C_1\) and \(C_2\), and hyperedges may either lie entirely within
one partition or be cut across the two. More generally, in this work we study
balanced \(k\)-way hypergraph partitioning, which seeks a partition of the node
set into \(k\) disjoint subsets of nearly equal size while minimizing a chosen
hypergraph cut objective. This problem is computationally challenging because,
in addition to minimizing the cut objective over all node assignments, it must
also enforce balance among the partition sizes. In contrast to the unbalanced
case, where polynomial-time solutions are available for certain two-way cut
objectives, balanced hypergraph partitioning is NP-hard even for
\(k=2\)~\cite{veldtHypergraphCutsGeneral2022}. Given the difficulty of finding
optimal balanced partitions, existing scalable methods typically rely on
spectral relaxation~\cite{liEfficientEffectiveAttributed2023}, greedy
Louvain-style methods~\cite{fengModularitybasedHypergraphClustering2023b},
local search~\cite{kabiljoSocialHashPartitioner2017b}, or ensembles of
heuristics~\cite{schlagHighQualityHypergraphPartitioning2023}.

Despite the success of these classical heuristics, finding high-quality partitions for increasingly large and dense hypergraphs remains a significant bottleneck due to the high-dimensional nature of the search space. Quantum computing  has emerged as a promising alternative for addressing such computationally intensive optimization tasks~\cite{abbas2024challenges}. Fundamental building blocks of quantum computers are qubits, which can exist in superpositions of basis states and become entangled with one another. By leveraging these uniquely quantum phenomena, certain problems can be solved with significantly lower complexity~\cite{grover1998advantages,shor1999polynomial,jozsa2003role}. 

Recent advances in hardware have brought quantum computers closer to utility-scale \cite{kim2023evidence, bluvstein2024logical,google2025quantum}.
Nevertheless, on the current Noisy Intermediate-Scale Quantum (NISQ) \cite{preskill2018quantum} devices, applications remain constrained by hardware limitations, including limited qubit counts, shallow circuit depths, and high error rates. While many quantum algorithms require fault-tolerant hardware, quantum annealing (QA) \cite{das2008colloquium,morita2008mathematical,hauke2020perspectives,mcgeoch2022adiabatic,rajak2023quantum} and variational quantum algorithms (VQAs) \cite{peruzzo2014variational,mcclean2016theory,jiang2017near,cerezo2021variational,tilly2022variational,bharti2022noisy,bravo2023variational,baaquie2023quantum} such as the Quantum Approximate Optimization Algorithm (QAOA) \cite{farhi2014quantum} offer a viable path for the NISQ era. These frameworks map combinatorial optimization problems onto quantum operators, which steer the evolution of the quantum state such that the optimal solution is encoded in the final measurement with high probability. Both QA and QAOA have been applied to various graph-optimization problems such as travelling salesman problem, maximum cut, and minimum k-way partitioning \cite{martovnak2004quantum,lucas2014ising,crooks2018performance, tse2018graph,cruz2019qubo,krauss2020solving, pramanik2020quantum,salehi2022unconstrained,abbassiQuantumApproaches2026}.  


In this paper, we study balanced hypergraph partitioning from the
perspective of quantum optimization, with formulations targeted at both QAOA
and quantum annealing. We derive and discuss binary optimization formulations
for a range of hypergraph cut objectives from the literature, covering both
two-way and multi-way settings. As a preliminary empirical validation, we focus
on the all-or-nothing cut on 3-uniform hypergraphs, where a quadratic formulation is
available, and evaluate the resulting formulation using simulated QAOA and
simulated annealing \cite{kirkpatrick1983optimization,van1987simulated,bertsimas1993simulated} against exact solutions. Our results provide initial
evidence that the proposed formulation is effective on small instances and
highlight the importance of the balance-penalty coefficient in trading off
feasibility and cut quality.


\section{Preliminaries}

\stitle{Quadratic Unconstrained Binary Optimization (QUBO)}
QUBO is a unifying framework for combinatorial optimization and a standard input
format for Ising-model-based solvers. It can represent diverse NP-hard
problems, such as partitioning, routing, and scheduling, in a common
mathematical form without explicit constraints. Instead, constraints are
absorbed into the objective as penalty terms, turning the problem into
energy minimization, where the global minimum corresponds to the optimum of the
original formulation.

Formally, a QUBO problem is defined as the minimization of a quadratic objective function:
\vspace{-1mm}
\begin{equation}
\label{eq:qubo}
\min_{x \in \{0,1\}^n} \; \xv^\mathsf{T} \QM \xv,
\end{equation}
where $\xv = (x_1,\dots,x_n)$ is a vector of $n$ binary variables and $\QM \in \mathbb{R}^{n \times n}$ is a real symmetric matrix. The diagonal entries $Q_{ii}$ represent the coefficients of linear terms, since $x_i^2 = x_i$, while the off-diagonal entries $Q_{ij}$ encode quadratic interactions between variables. Constructing an appropriate matrix $\QM$ is the central task in QUBO mapping, as it must balance the original objective against penalty terms that discourage the violation of problem constraints. 

A natural generalization of QUBO is Higher-order Unconstrained Binary Optimization (HUBO), in
which the objective may contain terms of degree greater than two. HUBO
objectives can often be handled directly in gate-based settings such as QAOA,
whereas QA typically requires a prior reduction to quadratic
form.

\stitle{Quantum optimization} QA and QAOA represent two of the most widely used paradigms for quantum optimization on NISQ-era hardware. Both approaches encode a combinatorial optimization problem into a problem Hamiltonian $H_P$, whose ground state corresponds to the optimal solution. Specifically, QUBO problems can be mapped to an Ising Hamiltonian 
\begin{equation}
    H_I= -\sum_i h_i \sigma^z_i - \sum_{i,j} J_{ij} \sigma^z_i \sigma^z_j,
\end{equation}
where $\sigma^z_i$ is a Pauli-z operator for the $i$th spin, $h_i$ represents the local longitudinal fields, and $J_{ij}$ denotes the pairwise coupling strengths. The mapping is done by transforming binary variables $x_i \in \{0, 1\}$ into spins $z_i \in \{-1, +1\}$ via the mapping $z_i = 2x_i - 1$, and setting $h_i=\QM_{ii}/2$ and $J_{ij}=\QM_{ij}/4$. The optimization task is thus equivalent to finding the ground state, the state that minimizes the expectation value
\begin{equation}
\label{eq:ising}
\min_{|\xv \rangle} \langle \xv | H_I|\xv\rangle = \min_{\mathbf{z} \in \{-1,+1\}^n} \left( -\sum_i h_i z_i - \sum_{i,j} J_{ij} z_i z_j \right).
\end{equation}
QA implements this optimization through a continuous adiabatic evolution of an Ising spin system. The system begins in the ground state $|   \psi _D\rangle$ of a driver Hamiltonian $H_D$ that does not commute with $H_P$ and is slowly evolved toward the problem Hamiltonian, ideally remaining in its ground state throughout  the transition. A typical choice for the driver Hamiltonian is $H_D= -\sum_i \sigma_i^x$, defined as the negative sum of Pauli-X operators acting on each qubit. Its ground state  is the uniform superposition of all $2^n$ computational basis states,
\begin{equation}
    |\psi_D \rangle=\frac{1}{\sqrt{2^n}} \sum_\xv |\xv \rangle.
\end{equation}
The total Hamiltonian of QA is given by 
    $H(t)=A(t) H_D + B(t) H_P$,
where $A(t)$ and $B(t)$ are function of time $t$ that control the transition from $H_P$ and to $H_D$. At the initial time $t=t_i$, $H(t_i)=H_D$, and at the final time, $H(t_f)=H_P$. As long as the Hamiltonian transition is sufficiently slow, the system's state $|\psi(t)\rangle$ closely follows the instantaneous ground state. Thus, the final state $|\psi(t_f)\rangle$ provides a close approximation of the ground state $|\psi_P\rangle$ of $H_P$.

    QAOA provides a digital, gate‑based approach. Under certain conditions, QAOA is a discrete approximation of QA \cite{boulebnane2025quantum}. However, QAOA can simulate the evolution of a broader class of Hamiltonians. The evolution is discretized into $L$ alternating applications of the problem Hamiltonian $H_P$ and a mixing Hamiltonian $H_M$. The state is initialzed as the ground state $|\psi_M\rangle$ of $H_M$, and its evolution is governed by variational parameters  $\boldsymbol{\beta}=(\beta_1,\dots,\beta_L) \in \mathbb{R}^L$ and $\boldsymbol{\gamma}=(\gamma_1,\dots,\gamma_L) \in \mathbb{R}^L$:
    \vspace{-4mm}
\begin{equation}
|\boldsymbol{\beta},\boldsymbol{\gamma} \rangle =\prod_{l=1}^L e^{-i\beta_l H_M}e^{-i\gamma_l H_P} | \psi_M \rangle.
\end{equation}
\vspace{-2mm}

 At each step of the classical optimization loop, the state $|\boldsymbol{\beta},\boldsymbol{\gamma} \rangle$ is measured in the computational basis to extract a bit-string $\mathbf{z}$. This result is used to evaluate the objective function and guide the classical optimizer in updating $\boldsymbol{\beta}$ and $\boldsymbol{\gamma}$. With increasing number of $K$ and optimization iterations, $|\boldsymbol{\beta},\boldsymbol{\gamma} \rangle$ converges to the ground state $|\psi_P\rangle$ of $H_P$.

Both QA and QAOA are probabilistic protocols, as the final state is generally a superposition of multiple computational basis states. Measuring this state yields a  single bit-string $z$ with a probability determined by the square of its corresponding state's amplitude. Consequently, the success of the protocol is assessed by the success probability, the likelihood of sampling a bit-string that represents an optimal or near-optimal solution. In practice, the algorithm is typically executed multiple times to ensure a high confidence of obtaining the correct outcome.

\stitle{Hypergraph}
A hypergraph is defined as \(H=(V,E)\), where \(V=\{v_1,\dots,v_n\}\) is the
set of \(n\) nodes and \(E=\{e_1,\dots,e_m\}\) is a collection of \(m\)
hyperedges, with each hyperedge \(e\in E\) being an arbitrary subset of \(V\),
i.e., \(E \subseteq 2^V\). Each hyperedge \(e\) is associated with a positive
weight \(w_e>0\); in the unweighted case, \(w_e=1\). The degree of a node
\(v\), denoted by \(d(v)\), is the number of hyperedges incident to \(v\). A
commonly studied special case is the \(r\)-uniform hypergraph, in which every
hyperedge has size \(r\).

\section{Problem Definition}
We study the balanced $k$-way hypergraph partitioning problem. The goal is to
assign each \emph{node} $v \in V$ to one of $k$ disjoint nonempty partitions 
$C=\{C_1,\dots,C_k\}$ such that
\(
\bigcup_{i=1}^k C_i = V
\)
and
\(
C_i \cap C_j = \varnothing
\)
for $i \neq j$, while keeping the partition sizes as even as possible.

\begin{figure}
    \centering
    \includegraphics[width=0.95\linewidth]{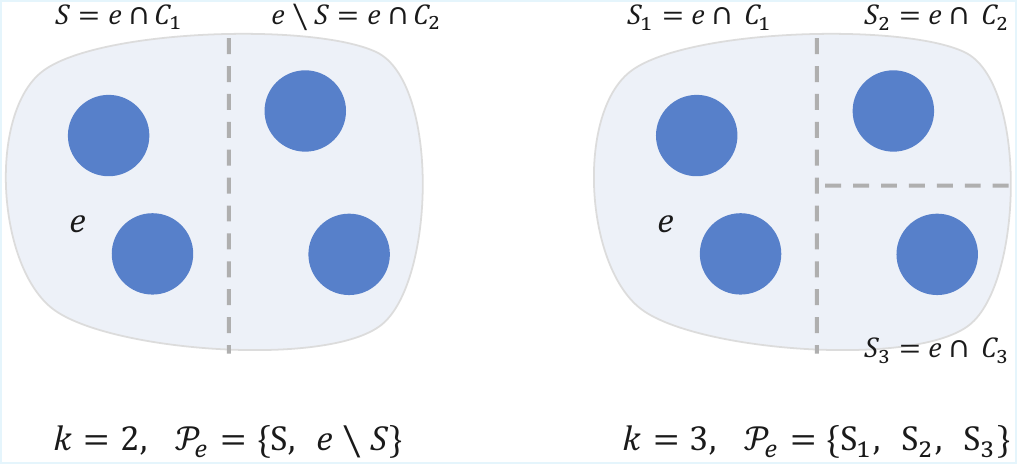}
    \vspace{-1mm}
    \caption{Illustration of the partition induced on hyperedge $e$.}
    \label{fig:hyperedge-cut}
    \vspace{-3mm}
\end{figure}

Each hyperedge $e$ inherits a partition from the node partition. In general,
its incident nodes are divided into disjoint nonempty subsets
$S_1,\dots,S_r$, where $1 \le r \le k$, such that $e=\bigcup_{i=1}^r S_i$.
We denote the partition of $e$ induced by the node partition $C$ as
$\mathcal P_e(C)=\{S_1,\dots,S_r\}$, or simply $\mathcal P_e$. For two-way
partitioning, this reduces to $\mathcal P_e=\{S,e\setminus S\}$ for some
$S \subseteq e$. Figure~\ref{fig:hyperedge-cut} illustrates these notations:
in the two-way case, the hyperedge is split into two parts,
$S=e\cap C_1$ and $e\setminus S=e\cap C_2$; in the multi-way case, the induced
partition consists of the nonempty intersections
with the global partitions, e.g., $\mathcal P_e=\{S_1,S_2,S_3\}$
for $k=3$. 

The contribution of hyperedge $e$ to the partitioning
objective is measured by a hyperedge cut function $f(\mathcal P_e)$, or
$f(S)$ in the two-way case.
Various choices of this cut function have
been studied in the literature; we refer readers to
\cite{veldtHypergraphCutsGeneral2022} for a review. A formal problem definition is as follows.

\begin{definition}[Balanced $k$-way hypergraph partitioning]
Given a weighted hypergraph $H=(V,E)$ with hyperedge weights $w_e$ and a
hyperedge cut function $f$, the balanced $k$-way hypergraph partitioning
problem is to find a partition $\mathcal C=\{C_1,\dots,C_k\}$ of $V$ that
minimizes
\begin{equation}
\label{eq:problem}
\begin{aligned}
\min_{\mathcal C=\{C_1,\dots,C_k\}} \quad
& \sum_{e \in E} w_e \, f\!\left(\mathcal P_e(\mathcal C)\right) \\
\text{s.t.} \quad
& \bigcup_{i=1}^k C_i = V, \quad
C_i \cap C_j = \varnothing \quad (i \neq j), \\
& C_i \neq \varnothing \quad \text{for all } i, \\
& \bigl||C_i|-|C_j|\bigr| \le 1
\quad \text{for all } i,j \in \{1,\dots,k\},
\end{aligned}
\end{equation}
where $\mathcal P_e(\mathcal C)$ denotes the partition of hyperedge $e$
induced by $\mathcal C$, i.e.,
\[
\mathcal P_e(\mathcal C)=\{\,e\cap C_i \mid e\cap C_i \neq \varnothing,\; i=1,\dots,k\,\}.
\]
\end{definition}
In this optimization problem, the first three constraints ensure a valid partition of the node set, i.e.,
each node belongs to exactly one partition and no partition is empty
(\emph{partition constraint}). The last constraint enforces balance by
requiring all partitions to have the same size when $k$ divides $n=|V|$, and
otherwise to differ in size by at most one (\emph{balance constraint}).

\stitle{Theoretical complexity}
The complexity of hypergraph partitioning depends strongly on both the number of
partitions and the choice of the cut function. For two-way partitioning ($k=2$)
without a balance constraint, cardinality-based submodular cut functions,
including several standard choices considered in this work, admit
polynomial-time solutions via reduction to graph cut problems~\cite{veldtHypergraphCutsGeneral2022}.
For multiway partitioning with $k>2$ and no balance constraint, the problem is
NP-hard when $k$ is part of the input, although an exact algorithm with running
time $n^{O(k)}$ has been given~\cite{chandrasekaranHypergraphKcutFixed2020a}.
Once balance constraints are imposed, however, hypergraph partitioning becomes
NP-hard for any $k\geq 2$ by reduction from the minimum graph bisection problem~\cite{gareySimplifiedNPcompleteProblems1974}.
Therefore, the problem studied in this work, namely balanced hypergraph
partitioning, is NP-hard.

\section{Objective Formulations}

For the hypergraph partitioning problem in Eq.~\ref{eq:problem}, the
corresponding binary optimization objective must encode both the cut cost and
the partitioning constraints. Depending on the chosen cut function, this
objective may take either a quadratic form, yielding a QUBO, or a higher-order
form, yielding a HUBO. We therefore decompose the objective into three
components:
\begin{equation}
\label{eq:overall-energy}
E(\xv)
= E_{\mathrm{cut}}(\xv)
+ \alpha \, E_{\mathrm{partition}}(\xv)
+ \lambda \, E_{\mathrm{balance}}(\xv),
\end{equation}
where \(E_{\mathrm{cut}}\) represents the chosen hyperedge cut objective,
\(E_{\mathrm{partition}}\) enforces that the binary assignment defines a valid
partition, and \(E_{\mathrm{balance}}\) enforces balanced partition sizes. The
coefficient \(\alpha>0\) controls the strength of the partition-validity
penalty, while \(\lambda>0\) is the balance-penalty coefficient.

In the following subsections, we treat two-way and multi-way partitioning
separately, since they require different objective constructions for encoding
partition assignments and constraints, and some cut objectives are specific to
one setting or take substantially different forms across the two. Accordingly,
we derive the corresponding forms of \(E_{\mathrm{cut}}\),
\(E_{\mathrm{partition}}\), and \(E_{\mathrm{balance}}\) for each case.

\subsection{Two-way Partitioning ($k=2$)}

For $k=2$, each vertex $v_i \in V$ is associated with a single binary variable
$x_i \in \{0,1\}$ indicating which partition it belongs to. We
interpret $x_i=1$ as assigning $v_i$ to the first partition and $x_i=0$ to the
second. The resulting QUBO therefore uses $n$ binary variables, i.e.,
$\xv \in \{0,1\}^n$.

Under this encoding, the validity constraints are satisfied implicitly: each
vertex is assigned to exactly one of the two partitions, so the two blocks are
automatically disjoint and their union is exactly $V$. Hence, no additional
validity penalty is required, and we set
\(E_{\mathrm{partition}}(\xv)=0\) and $E(\xv)
= E_{\mathrm{cut}}(\xv)
+ \lambda \, E_{\mathrm{balance}}(\xv)$.

For balanced partitioning, we use the constraint term
\begin{equation}
\label{eq:bi-constraint}
E_{\mathrm{balance}}(\xv)
= \left( \sum_{i=1}^n x_i - \frac{n}{2} \right)^2,
\end{equation}
which encourages the two partition sizes to be as equal as possible. This term
is minimized when the partitions have equal size if $n$ is even, and differ by
at most one if $n$ is odd. In particular, it guarantees that both partitions
are nonempty for $n \ge 2$.

For hypergraph partitioning without an explicit balance requirement,
nonemptiness is a weaker condition than the above balance constraint.
Nevertheless, the same penalty in Eq.~\eqref{eq:bi-constraint} can still be
retained with a smaller coefficient, so that it serves as a soft regularizer
against trivial collapsed assignments while allowing greater flexibility in
partition sizes. We therefore adopt Eq.~\eqref{eq:bi-constraint} as the
constraint term for two-way partitioning, with the penalty weight $\lambda$
chosen according to whether strict balance enforcement or only mild
regularization is desired.

We next discuss the cut term $E_{\mathrm{cut}}(\xv)$ for several hyperedge cut
models, beginning with the all-or-nothing cut. In the formulation, $S$ denotes the larger part of the hyperedge $e$ after the cut.

\stitle{All-or-nothing cut ($k=2$)} This cut function penalizes all hyperedge cuts by a uniform cost:
\begin{equation}
\label{eq-aon-cut}
    f_\text{AoN}(S)=\begin{cases}
0, & \text{if } S = e,\\
1, & \text{otherwise}.
\end{cases}
\end{equation}

An unconstrained binary optimization objective for the all-or-nothing cut is
given in~\cite{rodriguezQuantumAlgorithmsHypergraph2022} as
\begin{equation}
\label{eq-aon-hubo}
E_{\text{AoN}}(\xv) = \sum_{e \in E}
\left(
1 - \prod_{v_i \in e} x_i - \prod_{v_i \in e} (1-x_i)
\right).
\end{equation}
For each hyperedge $e$, the term $\prod_{v_i \in e} x_i$ equals $1$ only when
all vertices $v_i \in e$ are assigned to partition 1, while
$\prod_{v_i \in e} (1-x_i)$ equals $1$ only when all vertices $v_i \in e$ are
assigned to partition 0. Therefore, the expression contributes $0$ when all
vertices in $e$ lie in the same partition and $1$ otherwise, which matches the
all-or-nothing cut objective. 

For 3-uniform hypergraphs, Eq.~\eqref{eq-aon-hubo} reduces to a valid QUBO,
since the cubic terms from the two products cancel and the resulting
expression reduces to a quadratic polynomial. However, for hypergraphs with
larger hyperedges, Eq.~\eqref{eq-aon-hubo} becomes a HUBO rather than a QUBO.
This is not a
major issue for gate-based approaches such as QAOA, which can in principle
optimize higher-order objectives directly, although larger hyperedges generally
lead to deeper circuits. In contrast, quantum annealing requires a quadratic
form, so for \(|e|>3\) the objective must first be quadratized \cite{mandalCompressedQuadratizationHigher2020}, typically by
introducing auxiliary variables.

\stitle{Quadratic penalty} 
This cut function assigns a cost equal to the product of the sizes of the two
parts induced on a hyperedge:
\begin{equation}
    f_\mathrm{quadratic}(S)=|S|\cdot |e\setminus S|.
\end{equation}
For a hyperedge $e$, let $S=\{v_i\in e \mid x_i=1\}$. Then
$\sum_{v_i,v_j\in e} x_i(1-x_j)$ counts exactly the ordered pairs
$(v_i,v_j)$ such that $v_i\in S$ and $v_j\in e\setminus S$, and therefore equals
$|S|\cdot |e\setminus S|$. Summing over all hyperedges gives the QUBO objective
\begin{equation}
\label{eq:quadratic-qubo}
E_{\mathrm{quadratic}}(\xv)
= \sum_{e \in E} \sum_{v_i,v_j\in e} x_i(1-x_j).
\end{equation}

\stitle{Linear penalty}
This cut function penalizes a cut on hyperedge $e$ by the size of the smaller
side:
\begin{equation}
    f_\mathrm{linear}(S)=\min(|S|,|e\setminus S|).
\end{equation}
Let $s=|S|$ and $t=|e|$. Since $|e\setminus S|=t-s$, we can rewrite the penalty as
\begin{equation}
f_\mathrm{linear}(S)=\min(s,t-s)=\frac{t-|2s-t|}{2}
= -\left|s-\frac{t}{2}\right|+\frac{t}{2}.
\end{equation}
Because of the absolute value term, this function is not directly representable
as a QUBO.


\stitle{Hypergraph normalized cut~\cite{zhouLearningHypergraphsClustering2007}}
In the two-way setting, let $\{C_1,C_2\}$ denote the partition of the vertex set \(V\), and let \(S=e\cap C_1\), so that \(e\setminus S=e\cap C_2\). Then the contribution of hyperedge \(e\) can be written as
\begin{equation}
\label{eq-2-hncut}
f_{\mathrm{Ncut}}(S)
=
\left(
\frac{1}{\vol(C_1)}+\frac{1}{\vol(C_2)}
\right)
\frac{|S|\,|e\setminus S|}{|e|},
\end{equation}
where \(\vol(C)=\sum_{v_i\in C} d(v_i)\) denotes the volume of a vertex set \(C\).

The term
\(
|S|\,|e\setminus S|/|e|
\)
is a scaled version of the quadratic penalty. Therefore, if the normalization
factors \(1/\vol(C_1)\) and \(1/\vol(C_2)\) are omitted, the remaining objective
is equivalent in form to the quadratic penalty introduced above, up to the
hyperedge-dependent scaling factor \(1/|e|\), which can be absorbed into the
hyperedge weights.

However, the full normalized cut objective is not directly representable as a
QUBO or HUBO. Since \(\vol(C_1)\) and \(\vol(C_2)\) both depend on the binary assignment
variables, the resulting objective is a rational function. 

\subsection{Multi-way Partitioning ($k>2$)}
For $k>2$, each node $v_i$ is represented by a one-hot vector of $k$ binary variables $(x_{i1},\dots,x_{ik})$, where $x_{ic}=1$ indicates assignment to partition $c$. The QUBO thus uses $nk$ binary variables. The balance penalty is
\begin{equation}
\label{eq:k-balance}
E_{\mathrm{balance}}(\xv)
= \sum_{c=1}^k \left( \sum_{i=1}^n x_{ic} - \frac{n}{k} \right)^2,
\end{equation}
which enforces that all partitions have nearly equal size.

An additional constraint ensures that each node belongs to one partition only:
 \begin{equation}
     E_{\mathrm{partition}}(\xv)=\sum^n_{i=1} \left(\sum^k_{c=1}x_{ic}-1\right)^2
 \end{equation}

Let $\mathcal P(e)=\{S_1, \dots, S_r\}$ denote the nonempty parts cut from hyperedge $e$. We next discuss different multi-way cut objectives and their corresponding binary optimization formulations. 

\stitle{All-or-nothing cut}
Extending Eq.~\eqref{eq-aon-cut} and Eq.~\eqref{eq-aon-hubo} to the multi-way
setting, the all-or-nothing cut assigns a unit cost whenever a hyperedge spans
more than one partition, and zero otherwise. Formally,
\begin{equation}
\label{eq:aon-cut-multi}
f_{\mathrm{AoN}}(\mathcal P_e)=
\begin{cases}
0, & \text{if } |\mathcal P_e| = 1,\\
1, & \text{otherwise}.
\end{cases}
\end{equation}
That is, a hyperedge incurs no penalty if all of its vertices lie in the same
partition, and incurs a penalty of $1$ as soon as it is split.

Under the one-hot encoding, the corresponding HUBO objective is
\begin{equation}
\label{eq:multi-aon-hubo}
E_{\mathrm{AoN}}(\xv)
=
\sum_{e \in E}
\left(
1-\sum_{c=1}^k \prod_{v_i\in e} x_{ic}
\right),
\end{equation}
where \(\prod_{v_i\in e} x_{ic}\) equals $1$ if and only if all vertices in
hyperedge $e$ are assigned to partition $c$, and equals $0$ otherwise. Hence,
the inner summation is $1$ exactly when all vertices of $e$ belong to a single
partition, and $0$ otherwise, which matches the definition in
Eq.~\eqref{eq:aon-cut-multi}. 

\stitle{$\boldsymbol{K-1}$ penalty}
This cut function assigns a cost equal to the number of partitions spanned by a
hyperedge minus one:
\begin{equation}
    f_{K-1}(\mathcal P_e)=|\mathcal P_e|-1.
\end{equation}
Thus, a hyperedge incurs zero cost when all of its vertices lie in the same
partition, and the penalty increases linearly with the number of partitions it
touches. In the special case of two-way partitioning ($k=2$), this reduces to
the all-or-nothing cut.

Under the one-hot encoding, the corresponding HUBO objective can
be written as
\begin{equation}
\label{eq:kminusone-hobo}
E_{K-1}(\xv)
=
\sum_{e\in E}
\left(
\sum_{c=1}^k
\left(
1-\prod_{v_i\in e}(1-x_{ic})
\right)
-1
\right).
\end{equation}
For each hyperedge $e$ and partition $c$, the term
\(
1-\prod_{v_i\in e}(1-x_{ic})
\)
indicates whether $e$ has at least one vertex assigned to partition $c$.
Summing over all $c$ therefore counts the number of partitions spanned by $e$,
and subtracting $1$ yields the $K$-$1$ penalty.

\stitle{Hypergraph normalized cut}
The normalized cut objective can also be extended to the multi-way
setting~\cite{zhouLearningHypergraphsClustering2007,whangMEGAMultiviewSemisupervised2020b}:
\begin{equation}
\label{eq:k-hncut}
f_{\mathrm{Ncut}}(\mathcal P_e)
=
\sum_{i=1}^k
\frac{|e\cap C_i|\,|e\setminus C_i|}{\vol(C_i)\,|e|},
\end{equation}
where \(\vol(C_i)=\sum_{v_j\in C_i} d(v_j)\) denotes the volume of partition
\(C_i\).

As in the two-way case, the numerator is a scaled quadratic-type term, but the
denominator depends on the partition volumes. Substituting binary assignment
variables therefore yields a rational function rather than a 
polynomial. Hence, this cut function does not admit a direct QUBO or HUBO formulation.

Nevertheless, since this work focuses on balanced hypergraph partitioning, the
partition volumes \(\vol(C_i)\) are expected to be similar across clusters.
Under this assumption, the normalization factors \(1/\vol(C_i)\) may be viewed
as approximately constant, leading to the following unnormalized surrogate:
\begin{equation}
\label{eq:k-unnorm-quadratic}
f_{\mathrm{quadratic}}(\mathcal P_e)
=
\sum_{i=1}^k
\frac{|e\cap C_i|\,|e\setminus C_i|}{|e|},
\end{equation}
which is the natural multi-way extension of the quadratic penalty. Under one-hot
encoding, this yields the QUBO term
\begin{equation}
\label{eq:multi-quadratic-qubo}
E_{\mathrm{quadratic}}(\xv)
=
\sum_{e\in E}\sum_{c=1}^k
\frac{1}{|e|}
\sum_{v_i,v_j\in e}
x_{ic}(1-x_{jc}).
\end{equation}
Indeed, for each hyperedge \(e\) and partition \(C\), the inner summation counts
the ordered pairs \((v_i,v_j)\) such that \(v_i\in e\cap C\) and
\(v_j\in e\setminus C\), and is therefore equal to
\(|e\cap C|\,|e\setminus C|\). Thus, Eq.~\eqref{eq:multi-quadratic-qubo}
is exactly the QUBO corresponding to the unnormalized version of
Eq.~\eqref{eq:k-hncut}. When \(k=2\), this reduces to the two-way quadratic QUBO
in Eq.~\eqref{eq:quadratic-qubo}, up to the hyperedge-dependent scaling factor
\(1/|e|\).

\stitle{Hypergraph random walk conductance}
This recent hypergraph clustering objective is based on given transition probabilities for a random walk over the
hypergraph, with \(p(v_i,v_j)\) denoting the given probability that a random walk
starting from \(v_i\) ends at \(v_j\)~\cite{liEfficientEffectiveAttributed2023}.
For a partition \(\mathcal P=\{C_1,\dots,C_k\}\), the objective to be minimized is
\begin{equation}
\label{eq:hrwc-phi}
\Phi_{\mathrm{HRWC}}(\mathcal P)
=
\frac{1}{k}
\sum_{C\in\mathcal P}
\frac{1}{|C|}
\sum_{v_i\in C}
\sum_{v_j\notin C}
p(v_i,v_j).
\end{equation}

Under balanced \(k\)-way partitioning, we have \(|C|=n/k\) for every cluster
\(C\). Hence, the normalization factor \(1/|C|\) becomes a constant, and
Eq.~\eqref{eq:hrwc-phi} reduces to
\begin{equation}
\Phi_{\mathrm{HRWC}}(\mathcal P)
=
\frac{1}{n}
\sum_{C\in\mathcal P}
\sum_{v_i\in C}
\sum_{v_j\notin C}
p(v_i,v_j).
\end{equation}
Using one-hot variables \(x_{ic}\in\{0,1\}\), where \(x_{ic}=1\) indicates that
vertex \(v_i\) is assigned to partition \(C_c\), the corresponding QUBO term can
be written as
\begin{equation}
\label{eq:hrwc-qubo}
E_{\mathrm{HRWC}}(\xv)
=
\frac{1}{n}
\sum_{v_i,v_j\in V}
p(v_i,v_j)
\sum_{c=1}^k x_{ic}(1-x_{jc}).
\end{equation}
For any pair \((v_i,v_j)\), the inner summation
\(
\sum_{c=1}^k x_{ic}(1-x_{jc})
\)
equals \(1\) if \(v_i\) and \(v_j\) are assigned to different partitions, and
\(0\) otherwise. Thus, Eq.~\eqref{eq:hrwc-qubo} measures the total random-walk
escape probability across clusters. This formulation is also similar in form to
the multi-way quadratic QUBO in Eq.~\eqref{eq:multi-quadratic-qubo}; the key
difference is that here the coefficients are specified by the random-walk
transition probabilities \(p(v_i,v_j)\), rather than by hyperedge membership.
Since \(p(v_i,v_j)\) is treated as a given constant, Eq.~\eqref{eq:hrwc-qubo}
is a valid quadratic binary formulation.

\stitle{Alternative binary encoding}
Instead of one-hot encoding, a $k$-way partition can be represented using only
\(\lceil \log_2 k \rceil\) binary variables per vertex, by encoding each
partition index in binary. This reduces the number of variables from \(nk\) to
\(n\lceil \log_2 k \rceil\).

However, this saving comes at the cost of higher-order terms. Even identifying
whether a vertex belongs to a specific partition requires a term of degree
\(\lceil \log_2 k \rceil\). Consequently, both the balance term and the cut
objectives involve higher-order products, yielding HUBO formulations rather than
QUBOs. Such terms cannot be handled directly by quantum annealers without
quadratization, and for QAOA they generally imply deeper circuits and more
costly implementations.
Therefore, although binary encoding uses fewer qubits, we consider it less
practical than one-hot encoding for this work.

\section{Related Work}

\stitle{Hypergraph Partitioning}
Classical practical approaches include multilevel partitioners
for circuit and graph applications~\cite{karypisMultilevelHypergraphPartitioning1999a,schlagHighQualityHypergraphPartitioning2023},
local-search-based methods for large social networks~\cite{kabiljoSocialHashPartitioner2017b},
and modularity clustering methods~\cite{fengModularitybasedHypergraphClustering2023b,fengGraphRepresentationAttributed2025a}.
For hypergraph clustering, prior work has also generalized normalized cut and
related spectral objectives to hypergraphs and attributed hypergraphs, leading
to spectral or matrix-factorization-based algorithms~\cite{zhouLearningHypergraphsClustering2007,whangMEGAMultiviewSemisupervised2020b,liEfficientEffectiveAttributed2023}.
On the theoretical side, Veldt et al.~\cite{veldtHypergraphCutsGeneral2022}
provide a general treatment of hyperedge cut functions and the complexity of
the corresponding partitioning problems, while more recent work studies related
diffusion- and conductance-based objectives for local clustering~\cite{kamalLovaszSimonovitsTheoremHypergraphs2024a}.

\stitle{Quantum (Hyper)graph Partitioning}
Prior quantum work has focused mainly on graph partitioning, including quantum
annealing formulations for graph bisection and their multilevel
extensions~\cite{ushijima-mwesigwaGraphPartitioningUsing2017,ushijima-mwesigwaMultilevelCombinatorialOptimization2021},
QUBO formulations for constrained cut variants~\cite{cruzAQUBOFormulation2019,abbassiQuantumApproaches2026},
and QAOA-based approaches to application-specific settings such as image
segmentation~\cite{tseGraphCutSegmentation2019}. Other related directions
include multi-way graph partitioning with qudits~\cite{pramanikQuantumAssistedGraphClustering2021}
and learning-based selection of balance-penalty coefficients for minimum
bisection~\cite{rusnakovaQuantumAnnealingMinimum2025}. In contrast, quantum
algorithms for hypergraph partitioning remain largely unexplored. 
To our knowledge, the only directly related prior work is that of Rodriguez~\cite{rodriguezQuantumAlgorithmsHypergraph2022}, which provides a HUBO formulation for hypergraph two-way partitioning under the all-or-nothing objective. In contrast, we derive QUBO/HUBO formulations for both two-way and multi-way partitioning under general hyperedge cut functions.
We also note related quantum approaches on hypergraphs that are orthogonal to our focus, including quantum sparsification methods~\cite{liuQuantumSpeedupHypergraph2025} and theoretical analyses of QAOA on hypergraph-structured optimization problems~\cite{bassoPerformanceLimitationsQAOA2022}, which do not provide direct formulations for hypergraph partitioning.

\section{Experiments}
We present a preliminary experiment of balanced two-way
partitioning with the all-or-nothing cut on hypergraphs. Our evaluation focuses on this objective function for 3-uniform hypergraphs, since this
setting admits a direct QUBO formulation that can be easily simulated.
Restricting to small instances also enables exact verification via exhaustive
search. Code and data are available at \url{https://github.com/CyanideCentral/quantum-hypergraph-partitioning}.

\stitle{Settings}
We generate 100 random hypergraphs for each \(n \in \{8,\dots,15\}\) as our dataset. Each
hypergraph is unweighted, connected, and has uniform hyperedge size 3 and
average node degree \(\bar d \approx 5\). We compare four methods: exhaustive
search ({Exact}), KaHyPar~\cite{schlagHighQualityHypergraphPartitioning2023}, simulated QAOA ({sQAOA}) with \(X\)-mixer
Hamiltonian \(H_M=-\sum_i \sigma_i^x\), and simulated annealing ({SA}).
For {sQAOA}, we use the statevector simulator in Qiskit~\cite{qiskit2024}
with circuit depth \(1\), and select the best balanced partition among the
top-10 states with highest probability. For {SA}, we use the simulated
annealing sampler in Ocean SDK~\cite{dwave_ocean_sdk_docs} with 100 reads. For
both {sQAOA} and {SA}, we evaluate balance-penalty coefficients
\(\lambda \in \{0.3,1,3\}\). Each method is run 5 times with distinct
random seeds; we report averages and standard errors across these runs.

\stitle{Evaluation Protocol} 
We evaluate each solver on the all-or-nothing cut objective
(\(E_{\mathrm{AoN}}\) in Eq.~\eqref{eq-aon-hubo}), which is a valid QUBO for
size-3 hyperedges, over all hypergraph instances in the dataset. Since our target problem is
balanced partitioning, we report both \emph{feasibility rate}, defined as the
fraction of returned partitions that satisfy the balance constraint, and
\emph{optimality rate}, defined as the fraction of feasible partitions whose
all-or-nothing cut value matches the optimum computed by \textsc{Exact}. 

\stitle{Partitioning Performance}
\begin{figure}[t]
\centering

\begin{tikzpicture}
\begin{axis}[
    hide axis,
    xmin=0, xmax=1,
    ymin=0, ymax=1,
    legend columns=3,
    legend style={
        draw=none,
        fill=none,
        font=\small,
        column sep=0.35em
    }
]
\addlegendimage{blue, mark=o, mark options={solid, fill=white}, line width=1pt}
\addlegendentry{\textsc{sQAOA}, $\lambda=0.3$}
\addlegendimage{blue!70!black, mark=square*, mark options={solid, fill=white}, line width=1pt}
\addlegendentry{\textsc{sQAOA}, $\lambda=1$}
\addlegendimage{blue!40!black, mark=triangle*, mark options={solid, fill=white}, line width=1pt}
\addlegendentry{\textsc{sQAOA}, $\lambda=3$}
\addlegendimage{red, dashed, mark=o, mark options={solid, fill=white}, line width=1pt}
\addlegendentry{\textsc{SA}, $\lambda=0.3$}
\addlegendimage{red!70!black, dashed, mark=square*, mark options={solid, fill=white}, line width=1pt}
\addlegendentry{\textsc{SA}, $\lambda=1$}
\addlegendimage{red!40!black, dashed, mark=triangle*, mark options={solid, fill=white}, line width=1pt}
\addlegendentry{\textsc{SA}, $\lambda=3$}
\end{axis}
\end{tikzpicture}

\vspace{0.2em}

\begin{subfigure}[t]{\columnwidth}
\centering
\begin{tikzpicture}
\begin{axis}[
    width=\linewidth,
    height=0.52\linewidth,
    xlabel={$n$},
    xlabel style={at={(axis description cs:1.01,0)}, anchor=west},
    ylabel={Optimality rate},
    xmin=8, xmax=15,
    ymin=0, ymax=1.05,
    xtick={8,9,10,11,12,13,14,15},
    grid=both,
    grid style={gray!20},
    major grid style={gray!35},
    line width=1pt,
    mark size=2.2pt,
]

\addplot+[blue, mark=o, mark options={solid, fill=white}, error bars/.cd, y dir=both, y explicit] coordinates {
    (8,0.670000) +- (0,0.147596)
    (9,0.553602) +- (0,0.191733)
    (10,0.537037) +- (0,0.208364)
    (11,0.491573) +- (0,0.200516)
    (12,0.538121) +- (0,0.165671)
    (13,0.457814) +- (0,0.158182)
    (14,0.375685) +- (0,0.149790)
    (15,0.366608) +- (0,0.119754)
};
\addplot+[blue!80!black, mark=square*, mark options={solid, fill=white}, error bars/.cd, y dir=both, y explicit] coordinates {
    (8,0.863869) +- (0,0.107969)
    (9,0.853980) +- (0,0.126126)
    (10,0.783441) +- (0,0.130669)
    (11,0.666000) +- (0,0.145481)
    (12,0.414209) +- (0,0.118821)
    (13,0.370000) +- (0,0.121655)
    (14,0.183161) +- (0,0.060193)
    (15,0.102000) +- (0,0.041415)
};
\addplot+[blue!65!black, mark=triangle*, mark options={solid, fill=white}, error bars/.cd, y dir=both, y explicit] coordinates {
    (8,0.437557) +- (0,0.102900)
    (9,0.262000) +- (0,0.079468)
    (10,0.202649) +- (0,0.082534)
    (11,0.104000) +- (0,0.028085)
    (12,0.087444) +- (0,0.052396)
    (13,0.048000) +- (0,0.020080)
    (14,0.035448) +- (0,0.015771)
    (15,0.010000) +- (0,0.002829)
};
\addplot+[red, dashed, mark=o, mark options={solid, fill=white}, error bars/.cd, y dir=both, y explicit] coordinates {
    (8,0.000000) +- (0,0.000000)
    (9,1.000000) +- (0,0.000000)
    (10,0.000000) +- (0,0.000000)
    (11,1.000000) +- (0,0.000000)
    (12,1.000000) +- (0,0.000000)
    (13,1.000000) +- (0,0.000000)
    (14,1.000000) +- (0,0.000000)
    (15,1.000000) +- (0,0.000000)
};
\addplot+[red!80!black, dashed, mark=square*, mark options={solid, fill=white}, error bars/.cd, y dir=both, y explicit] coordinates {
    (8,1.000000) +- (0,0.000000)
    (9,1.000000) +- (0,0.000000)
    (10,1.000000) +- (0,0.000000)
    (11,1.000000) +- (0,0.000000)
    (12,1.000000) +- (0,0.000000)
    (13,1.000000) +- (0,0.000000)
    (14,1.000000) +- (0,0.000000)
    (15,1.000000) +- (0,0.000000)
};
\addplot+[red!65!black, dashed, mark=triangle*, mark options={solid, fill=white}, error bars/.cd, y dir=both, y explicit] coordinates {
    (8,1.000000) +- (0,0.000000)
    (9,1.000000) +- (0,0.000000)
    (10,1.000000) +- (0,0.000000)
    (11,1.000000) +- (0,0.000000)
    (12,1.000000) +- (0,0.000000)
    (13,1.000000) +- (0,0.000000)
    (14,1.000000) +- (0,0.000000)
    (15,1.000000) +- (0,0.000000)
};

\end{axis}
\end{tikzpicture}
\caption{Optimality rate (Exact $=1$).}
\end{subfigure}

\vspace{0.45em}

\begin{subfigure}[t]{\columnwidth}
\centering
\begin{tikzpicture}
\begin{axis}[
    width=\linewidth,
    height=0.52\linewidth,
    xlabel={$n$},
    xlabel style={at={(axis description cs:1.01,0)}, anchor=west},
    ylabel={Feasibility rate},
    xmin=8, xmax=15,
    ymin=0, ymax=1.05,
    xtick={8,9,10,11,12,13,14,15},
    grid=both,
    grid style={gray!20},
    major grid style={gray!35},
    line width=1pt,
    mark size=2.2pt,
]

\addplot+[blue, mark=o, mark options={solid, fill=white}, error bars/.cd, y dir=both, y explicit] coordinates {
    (8,0.288000) +- (0,0.121372)
    (9,0.370000) +- (0,0.142070)
    (10,0.262000) +- (0,0.132406)
    (11,0.422000) +- (0,0.154023)
    (12,0.332000) +- (0,0.131800)
    (13,0.498000) +- (0,0.170585)
    (14,0.432000) +- (0,0.148146)
    (15,0.566000) +- (0,0.174736)
};
\addplot+[blue!80!black, mark=square*, mark options={solid, fill=white}, error bars/.cd, y dir=both, y explicit] coordinates {
    (8,0.960000) +- (0,0.009381)
    (9,0.996000) +- (0,0.002191)
    (10,0.988000) +- (0,0.004382)
    (11,1.000000) +- (0,0.000000)
    (12,0.980000) +- (0,0.011314)
    (13,1.000000) +- (0,0.000000)
    (14,0.962000) +- (0,0.023732)
    (15,0.998000) +- (0,0.001789)
};
\addplot+[blue!65!black, mark=triangle*, mark options={solid, fill=white}, error bars/.cd, y dir=both, y explicit] coordinates {
    (8,0.974000) +- (0,0.008764)
    (9,1.000000) +- (0,0.000000)
    (10,0.974000) +- (0,0.009633)
    (11,1.000000) +- (0,0.000000)
    (12,0.976000) +- (0,0.008294)
    (13,1.000000) +- (0,0.000000)
    (14,0.962000) +- (0,0.011454)
    (15,0.996000) +- (0,0.003578)
};
\addplot+[red, dashed, mark=o, mark options={solid, fill=white}, error bars/.cd, y dir=both, y explicit] coordinates {
    (8,0.000000) +- (0,0.000000)
    (9,0.010000) +- (0,0.000000)
    (10,0.000000) +- (0,0.000000)
    (11,0.014000) +- (0,0.002191)
    (12,0.020000) +- (0,0.000000)
    (13,0.098000) +- (0,0.003346)
    (14,0.120000) +- (0,0.000000)
    (15,0.110000) +- (0,0.000000)
};
\addplot+[red!80!black, dashed, mark=square*, mark options={solid, fill=white}, error bars/.cd, y dir=both, y explicit] coordinates {
    (8,0.378000) +- (0,0.008672)
    (9,0.786000) +- (0,0.010040)
    (10,0.486000) +- (0,0.007798)
    (11,0.864000) +- (0,0.014311)
    (12,0.582000) +- (0,0.015849)
    (13,0.898000) +- (0,0.013682)
    (14,0.618000) +- (0,0.011099)
    (15,0.930000) +- (0,0.006324)
};
\addplot+[red!65!black, dashed, mark=triangle*, mark options={solid, fill=white}, error bars/.cd, y dir=both, y explicit] coordinates {
    (8,0.994000) +- (0,0.003578)
    (9,1.000000) +- (0,0.000000)
    (10,0.998000) +- (0,0.001789)
    (11,1.000000) +- (0,0.000000)
    (12,1.000000) +- (0,0.000000)
    (13,1.000000) +- (0,0.000000)
    (14,1.000000) +- (0,0.000000)
    (15,1.000000) +- (0,0.000000)
};

\end{axis}
\end{tikzpicture}
\caption{Feasibility rate (Exact $=1$).}
\end{subfigure}
\vspace{-1mm}
\caption{Preliminary results on 3-uniform hypergraphs with different sizes. Error bars show standard error across five runs.}
\label{fig:exp-opt-feas}
\vspace{-3mm}
\end{figure}
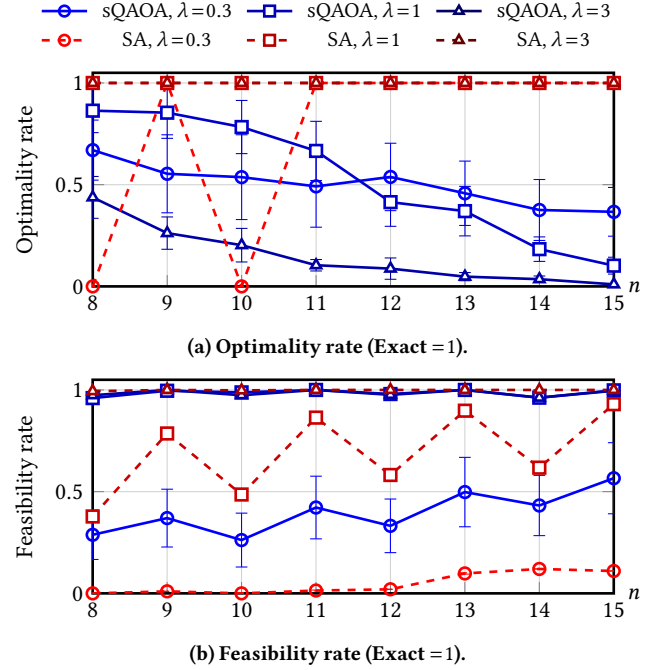
Figure~\ref{fig:exp-opt-feas} shows the feasibility and optimality rates across
different problem sizes and balance-penalties. KaHyPar attains the optimal balanced all-or-nothing cut on all tested instances, matching Exact in this small-scale setting. 

\textsc{sQAOA} with \(\lambda=1\) performs well on smaller instances, producing
optimal balanced partitions for over 90\% of the cases with \(n<10\). However,
its performance degrades as the hypergraph size increases. With
\(\lambda=0.3\), \textsc{sQAOA} maintains a moderate optimality rate but often violates the balance constraint. The large variation across
runs suggests that \textsc{sQAOA} is sensitive to random initialization.

In contrast, \textsc{SA} with \(\lambda=3\) consistently returns optimal
balanced partitions for almost all tested instances. Reducing the penalty weight to
\(\lambda=1\) lowers the feasibility rate, although the returned solutions
remain optimal whenever they are feasible. With \(\lambda=0.3\), \textsc{SA}
rarely produces balanced partitions. \textsc{SA} also shows greater robustness across random seeds, with only negligible standard errors for both metrics.

\stitle{Effect of the Balance Penalty} The coefficient \(\lambda\) in Eq.~\eqref{eq:overall-energy} controls the
tradeoff between cut minimization and balance enforcement. When \(\lambda\) is
too small, the solver may favor lower-cut but imbalanced solutions, as seen for
\textsc{SA} with \(\lambda=0.3\), which usually fails to satisfy the balance
constraint. On the other hand, overly large \(\lambda\) may overemphasize the
balance term and reduce solution quality, as illustrated by the sharp drop in
the optimality rate of \textsc{sQAOA} when \(\lambda\) is increased to 3.

Overall, these results suggest that moderate values of \(\lambda\) provide the
best tradeoff for this formulation, although the preferred range may depend on
both the solver and the instance distribution. 

\section{Conclusions}
We studied balanced hypergraph partitioning as a target problem for quantum
optimization. In particular, we formalized the problem under general
hyperedge cut functions and derived corresponding binary optimization formulations for both
two-way and multi-way partitioning. The analysis shows that some widely used
objectives, such as quadratic-style cut functions, admit direct QUBO
encodings, while others require higher-order formulations or approximations.

As a preliminary evaluation, we focused on balanced two-way partitioning with
the all-or-nothing cut on 3-uniform hypergraphs, where the objective can be
encoded directly as a QUBO. Experiments on small instances confirmed that the
formulation can recover optimal balanced partitions under suitable solver and
penalty settings, and also highlighted the importance of the balance-penalty
coefficient in controlling the tradeoff between feasibility and cut quality.

Future work includes extending to multi-way partitioning, evaluating larger and more diverse hypergraphs, real-hardware experiments, and systematic strategies
for choosing penalty coefficients.

\begin{acks}
This research has received funding from the research project entitled “Quantum Software Consortium: Exploring Distributed Quantum Solutions for Canada” (QSC). QSC is financed by the National Sciences and Engineering Research Council of Canada (NSERC) Alliance Consortia Quantum program under grant number ALLRP587590-23.
\end{acks}

\bibliographystyle{ACM-Reference-Format}
\bibliography{refs}

\appendix

\end{document}